\documentclass{ieeeaccess}
\usepackage{cite}
\usepackage{amsmath,amssymb,amsfonts}
\usepackage{algorithmic}

\usepackage{graphicx,color,overpic}
\usepackage{subfigure}  
\usepackage{caption,setspace}
\usepackage{float}
\usepackage{enumerate}\textbf{}
\usepackage{enumitem}
\graphicspath{{figures-pdf/}}

\usepackage{textcomp}
\def\BibTeX{{\rm B\kern-.05em{\sc i\kern-.025em b}\kern-.08em
    T\kern-.1667em\lower.7ex\hbox{E}\kern-.125emX}}

\usepackage{hyperref}

\newtheorem{fact}{Fact}

\let\oldbibitem\bibitem
\def\bibitem{\vfill\oldbibitem}

\usepackage[normalem]{ulem}

\newlength\figsep
\newlength\twoimagewidth
\newlength\threeimagewidth
\newlength\BigOneImW
\newlength\imagewidth
\newlength\DoubleThreeImW
\begin{document}
	
\history{Date of publication xxxx 00, 0000, date of current version xxxx 00, 0000.}
\doi{10.1109/ACCESS.2018.2883690}

\title{Cryptanalysis of a Chaotic Image Encryption Algorithm Based on Information Entropy}

\author{\uppercase{Chengqing Li}\authorrefmark{1}, \uppercase{Dongdong Lin}\authorrefmark{1}, \uppercase{Bingbing Feng}\authorrefmark{2},  \uppercase{Jinhu L\"u}\authorrefmark{3}, and Feng Hao\authorrefmark{4}}

\address[1]{School of Computer Science and Electronic Engineering,\\
	Hunan University, Changsha 410082, China}
\address[2]{College of Information Engineering,\\
	Xiangtan University, Xiangtan 411105, Hunan, China}
\address[3]{School of Automation Science and Electrical Engineering, \\	
	Beihang University, Beijing 100083, China}
\address[4]{Department of Computer Science, University of Warwick, Coventry CV4~7AL, UK}

\tfootnote{This research was supported by the Natural Science Foundation of China (No.~61772447, 61532020, U1736113).}

\markboth
{C. Li \headeretal: Cryptanalysis of an Image Encryption Algorithm}
{C. Li \headeretal: Cryptanalysis of an Image Encryption Algorithm}

\corresp{Corresponding author: C Li (e-mail: chengqingg@qq.com).}

\begin{abstract}
Recently, a chaotic image encryption algorithm based on information entropy (IEAIE) was proposed.
This paper scrutinizes the security properties of the algorithm and evaluates the validity of the used quantifiable security metrics. When the round number is only one, the equivalent secret key of every basic operation of IEAIE can be recovered with a differential attack separately. Some common insecurity problems in the field of chaotic image encryption are found in IEAIE, e.g. the short orbits of the digital chaotic system and the invalid sensitivity mechanism built on information entropy of the plain image. Even worse, each security metric is questionable, which undermines the security credibility of IEAIE. Hence, IEAIE can only serve as a counterexample for illustrating common pitfalls in designing secure communication method for image data.
\end{abstract}
\begin{keywords}
Chaotic cryptanalysis, multimedia cryptography, image encryption, secure communication, privacy protection.
\end{keywords}

\titlepgskip=-15pt
\maketitle

\section{Introduction}

With the popularity of imaging sensors in smartphones and various video recording scenarios, e.g. dashboard camera and closed-circuit television (CCTV),
a vast volume of multimedia data are recorded every day \cite{KunWang:2017selective,du2018big:IEEECM}. Meanwhile, the fast network transmission
technique allows them to be transmitted among cloud servers, social media platforms, and personal cellphones with ever-growing speed and scope.
Once a multimedia file containing some personal privacy information leaves the original control scope, they may threaten the owner and the related persons very quickly. So,  the security and privacy of multimedia data have become the concerns of everyone living in the cyberspace. To respond to such a challenge, a large number of multimedia privacy protections and preservation schemes were proposed in the past two decades \cite{Zhangxing:privacy:Access2018,Chatterjee:chebyshev:TDSC2018}.

One of the well-known features of chaos is the so-called butterfly effect: if a butterfly flips its wings
in Brazil, tomorrow Texas, USA will have a storm. In a more scientific term, we say a
system is very sensitive to the initial condition, i.e., a small change at the very
beginning will eventually lead to a completely different result. This implies
unpredictability because an accurate measurement of the initial condition is in
principle impossible. As the sensitivity and unpredictability are some good features we want to have in applications like secure communications and (pseudo-) random number generation, many researchers around the world have tried to apply chaos to build various cryptographic primitives:
permutation relation \cite{Cqli:Scramble:IM17}, pseudo-random number generator \cite{Hua:sine:IS2016,Hua2018SP}, hash function \cite{lin2017hash:IJBC}, private-key encryption scheme \cite{zhu2012novel:OC12,Shen:IJBC:2017}, public-key encryption scheme \cite{shakiba2016chebyshev:IJBC}, authentication \cite{Chatterjee:chebyshev:TDSC2018}, secure communication based on synchronization \cite{abd2017synchronization:ND}, secret-key share (agreement) algorithm \cite{zhang2017privacy:JBHI}, data hiding \cite{Ge:IJBC:2016}, and privacy protection \cite{Sun:IJBC:2017}. The main objective of chaotic cryptanalysis is to disclose the information about the secret key of a chaotic encryption (or secure communication) scheme under all kinds of security models: \textit{ciphertext-only attack} \cite{Cqli:Scramble:IM17}, \textit{known-plaintext attack} \cite{Li:logistic:ND2014,Cqli:hierarchical:SP2016}, \textit{chosen-plaintext attack} \cite{Li:RCES:JSS2008,cqli:autoblock:IEEEM18}, \textit{chosen-ciphertext attack} \cite{ge2017cryptanalyzing:ND}, and \textit{impossible differential attack} \cite{yap2015cryptanalysis:ND}. Meanwhile, chaotic cryptanalysis also provides a novel perspective to study the dynamical properties of the underlying chaotic system. As degradation of any chaotic system definitely happens in a digital domain \cite{WangQX:HDDCS:TCAS2016,luo:perturbation:IJBC2018}, a chaos-based encryption scheme may own some special security defects that do not exist in the non-chaotic encryption schemes \cite{Alvarez:IJBC:2006,ozkaynak:review:ND18,Zhucongxu:Tent:Access18}.

In \cite{yegd:IJBC:2018}, a chaotic image encryption algorithm was proposed using information entropy value calculated from the plain-image, which is named as IEAIE in this paper. In the algorithm, a pseudo-random number sequence generated by the two-dimensional Logistic-adjusted-Sine map (2D-LASM) proposed in \cite{Hua:sine:IS2016} is used to control a combination of some basic operations, including position permutation and modulo addition. Especially, the information entropy of the plain-image is used to build up a sensitivity mechanism of the encryption result of IEAIE on the plain-image. This paper reports security defects of the chaos-based pseudo-random number generator and the sensitivity mechanism. As for one round version of IEAIE, its three basic parts can be broken with a strategy of the divide-and-conquer technique in the scenario of differential attack. In addition, each used security metric is questioned from the perspective of modern cryptanalysis.

The rest of the paper is organized as follows. Section~\ref{algorithm} briefly introduces the algorithm IEAIE. Section~\ref{cryptanalysis} presents cryptanalysis of IEAIE by disproving security metrics used for IEAIE. The last section concludes the paper.

\section{Concise description of IEAIE}
\label{algorithm}

IEAIE ignores any special storage format of image data and just treats it as text data, which is represented as a $M\times N$ 8-bit matrix $\mathbf{I}$ \footnote{The transform (5) in \cite{yegd:IJBC:2018} can not always generate bijective (one-to-one) permutation mapping and should be corrected to assure successful decryption of IEAIE}.
	
\begin{itemize}
	
	\item\textit{The secret key} is composed of two sets of initial conditions of
	2D-LASM
	\begin{equation}
		\begin{cases}
			x_{i+1}  =  \sin( \pi\cdot \mu\cdot (y_i+3)\cdot x_i\cdot(1-x_i)),\\
			y_{i+1}  =  \sin( \pi\cdot \mu\cdot (x_i+3)\cdot y_i\cdot(1-y_i)),
		\end{cases}
		\label{eq:2DLASM}
	\end{equation}
	$(x_0, y_0)$ and $(x'_0, y'_0)$, where $\mu \in [0.37, 0.38]\cup [0.4, 0.42] \cup [0.44, 0.93]$.
	
	\item\textit{Keystream generation procedure}:
	1) iterate 2D-LASM \eqref{eq:2DLASM} from initial condition
	\begin{equation}
		\begin{cases}
			\bar{x}_0 = ( x_0 + \frac{s+1}{s+x'_0+y'_0+1}) \bmod 1\\
			\bar{y}_0 = ( y_0 + \frac{s+2}{s+x'_0+y'_0+2}) \bmod 1
		\end{cases}
		\label{eq:update}
	\end{equation}	
	$200+\frac{M\cdot N}{2}$ times, and from the 201-th iteration, assign the obtained sequence into an $M\times N$ matrix $\mathbf{P}$ in the raster order, where
\begin{equation}
s=H( \mathbf{I} ),	
\label{eq:entropys}
\end{equation}
$H(\mathbf{X})$ is the information entropy value of image block $\mathbf{X}$, namely
	\begin{equation}
		H(\mathbf{X})=-\sum_{i=0}^{2^8-1}p(\phi_i) \cdot \log_2(p(\phi_i)),
		\label{eq:entropy}
	\end{equation}
	$\phi_i$ is the pixel of value $i$ in $\mathbf{X}$, and $p(\phi_i)$ denotes the ratio between the number of $\phi_i$ in $\mathbf{X}$ and $\mathit{M\cdot N}$. In this paper, $x\bmod n=x-n\lfloor x/n \rfloor$, where
	$\lfloor \cdot \rfloor$ denotes the floor function.
	
	3) set
	\begin{equation}
		\begin{cases}
			\mathbf{u}= \lceil \mathbf{u}'\cdot 10^{14} \rceil \bmod M +1, \\
			\mathbf{v}= \lceil \mathbf{v}'\cdot 10^{14} \rceil \bmod N +1,
		\end{cases}
		\label{eq:getMatrixP}
	\end{equation}
	where $\mathbf{u}'$ is the $a$-th row of $\mathbf{P}$, $\mathbf{v}'$ is the $b$-th column of $\mathbf{P}$
	(scalar multiplication and addition are performed if a matrix or vector is involved, the same hereinafter),
	\begin{equation}
		\begin{cases}
			a= \lceil (x_0+y_0+1)\cdot 10^7 \rceil \bmod M+1, \\
			b= \lceil (x'_0+y'_0+2)\cdot 10^7 \rceil \bmod N+1,
		\end{cases}
		\label{convert2}
	\end{equation}
	and $\lceil \cdot \rceil$ denotes the ceil function. Separately conduct the two vectors $\mathbf{u}$ and $\mathbf{v}$ with the following way:
	if there are elements of the same value, change one as the least number that does not exist in the updated vector.
	
	3) iterate 2D-LASM \eqref{eq:2DLASM} from initial condition
	\begin{equation}
		\begin{cases}
			\bar{x}'_0 = ( x'_0 + \frac{1}{x_0+y_0+1}) \bmod 1\\
			\bar{y}'_0 = ( y'_0 + \frac{2}{x_0+y_0+2}) \bmod 1
		\end{cases}
		\label{eq:getMatrixK}
	\end{equation}
	$200+\frac{M\cdot N}{2}$ times; starting from the 201-th iteration, transform every element of the generated sequence by
	\begin{equation}
		f(x)=\lceil x\cdot 10^{14} \rceil \bmod 256
		\label{convert}
	\end{equation}
	and set the results into an $M\times N$ matrix $\mathbf{K}$ in the raster order.
	
	\item\textit{Encryption procedure}:
	
\begin{itemize}
		\item \textit{Horizontal permutation}:	
		for $j=1\sim N$, move the $j$-th column of $\mathbf{I}$ to the $\mathbf{u}(j)$-th one of
		$\mathbf{B}^*$, namely $\mathbf{B}^*(:, \mathbf{u}(j))=\mathbf{I}(:, j)$.
		
		\item \textit{Vertical permutation}:
		for $i=1\sim M$, move the $i$-th row of $\mathbf{B}^*$ to the $\mathbf{v}(i)$ row of $\mathbf{B}$,
		i.e. $\mathbf{B}(\mathbf{v}(i), :)=\mathbf{B}^*(i, :)$.
		
		\item \textit{Changing gray distribution with a constant matrix $\mathbf{T}$}: for $i=1\sim M$, $j=1\sim N$, do
		\begin{equation}\label{eq:changeDistribution}
			\mathbf{R}(i, j)=(\mathbf{B}(i, j)+\mathbf{T}(i, j))\bmod 256,
		\end{equation}
		where $\mathbf{T}(i, j)=M\cdot N+ i+j$.
		
		\item \textit{Diffusion encryption:} for $i=1\sim M$, $j=1\sim N$, set
		\begin{align}
			\mathbf{C}(i, j)=( \mathbf{R}(i, j)+d_j\cdot \mathbf{C}(i, j-1)+ \nonumber\\  d_j\cdot \mathbf{K}(i, j)+ \mathbf{K}(i, d_j)) \bmod 256,
			\label{eq:diffusion}
		\end{align}
		where $\mathbf{C}(i, \mathit{N}+1)=\mathbf{C}(i, 0)\equiv 0$,
		\begin{equation}
		d_j=\lceil H( \mathbf{R}_j )\cdot 10^{14} \rceil \bmod N+1,	
		\label{parameterD}
		\end{equation}
	and $\mathbf{R}_j=\{\mathbf{R}(i, k)\}_{i=1, k=j+1}^{M, N}$.
\end{itemize}
	
	\item \textit{Repeation:} Repeat the above four steps one more round.
	
	\item \textit{Decryption procedure} is similar to the encryption one except the following points:
	1) the order of the above four main steps is reversed; 2) every operation in each main step is replaced by its inverse version.
\end{itemize}

The horizontal and vertical permutations on the plain-image controlled by $\mathbf{u}$ and $\mathbf{v}$ can be equivalently represented by a permutation matrix  $\mathbf{P}$ as \cite{Cqli:hierarchical:SP2016}, namely
\begin{equation}
\mathbf{B}( \mathbf{P}(i, j ))=\mathbf{I}(i, j),
\label{equivalentPermut}
\end{equation}
where $i=1\sim M$, and $j=1\sim N$. 

\begin{figure*}[!htb]
	\centering
	\begin{minipage}{\threeimagewidth}
		\centering
		\includegraphics[width=\threeimagewidth]{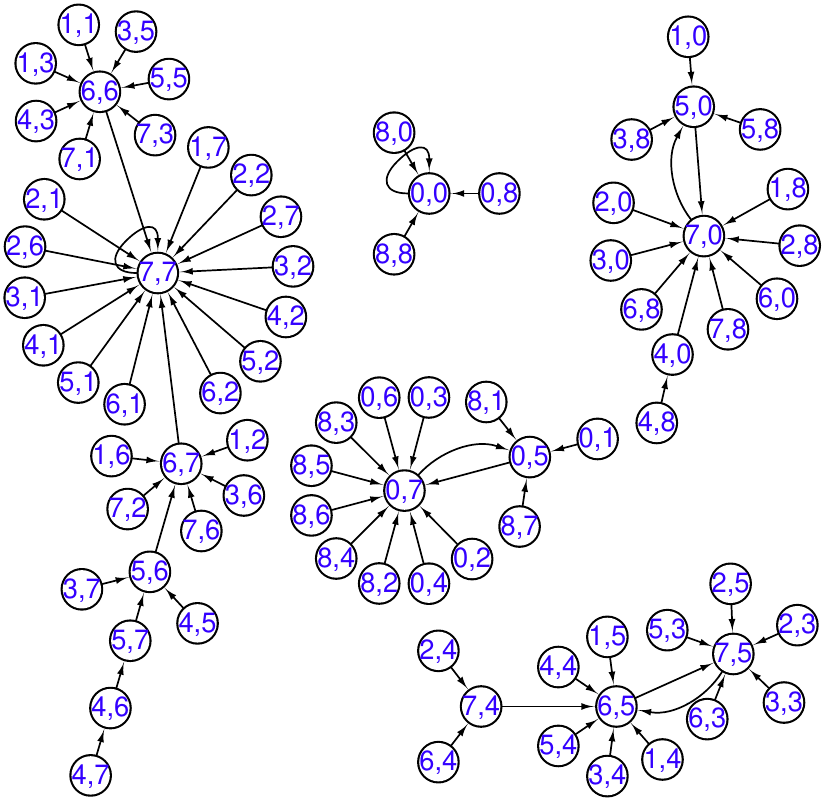}
		a)
	\end{minipage} \hspace{\figsep}
	\begin{minipage}{\threeimagewidth}
		\centering
		\includegraphics[width=\threeimagewidth]{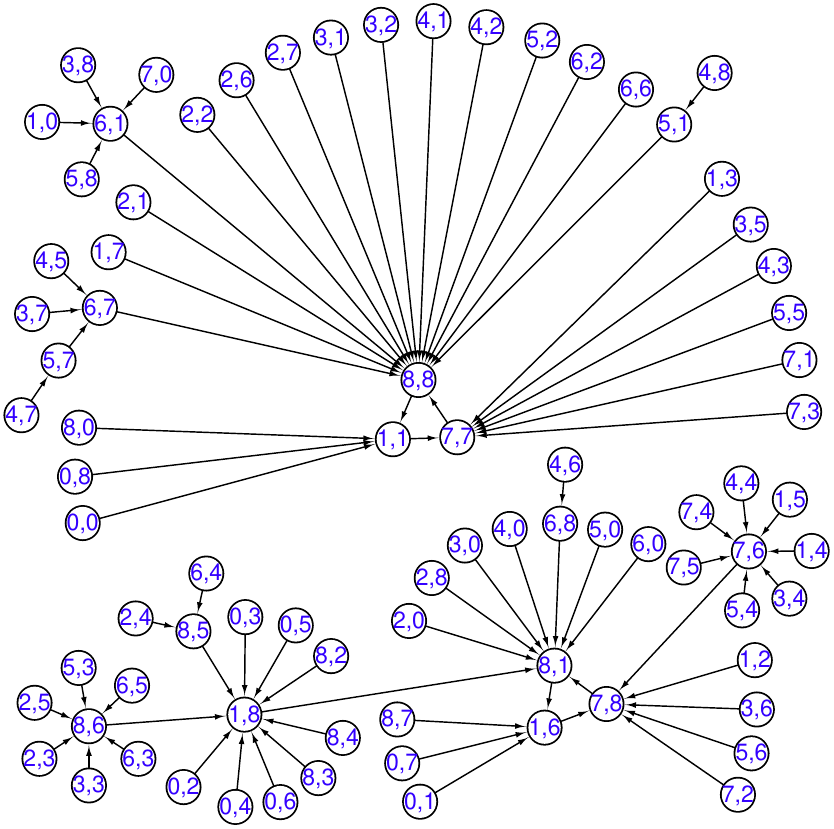}
		b)
	\end{minipage}\hspace{\figsep}
	\begin{minipage}{\threeimagewidth}
		\centering
		\includegraphics[width=\threeimagewidth]{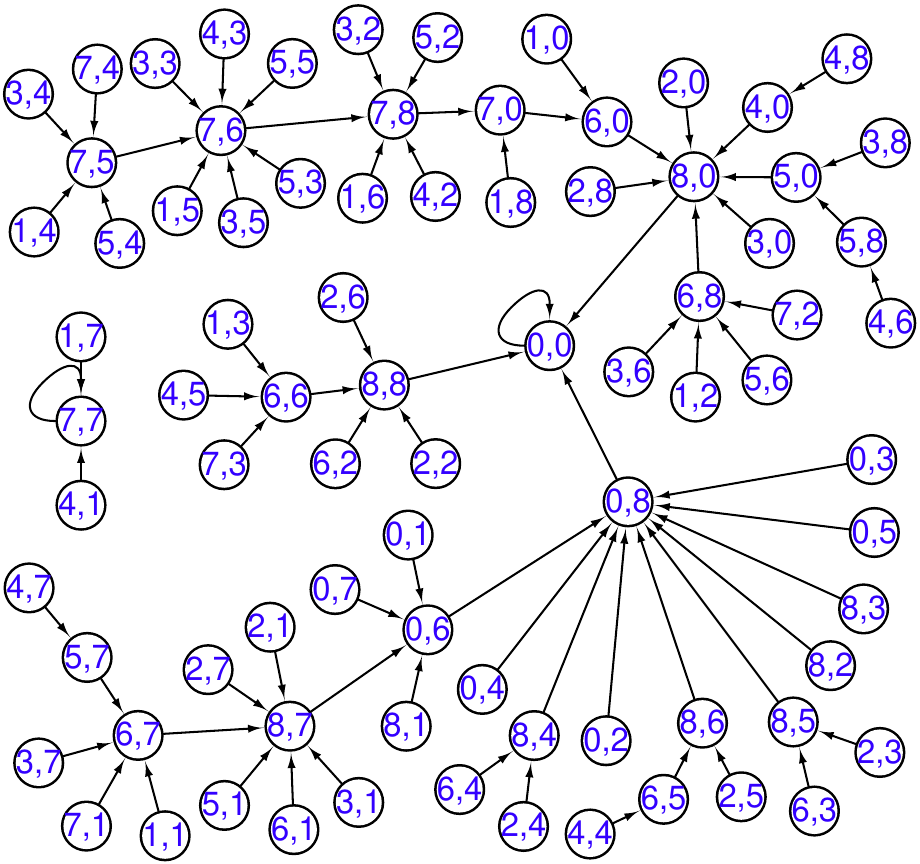}
		c)
	\end{minipage}
	\caption{The functional graph of 2D-LASM under 3-bit fixed-point precision for different quantization strategies: a) floor; b) round; c) ceil, where the pair of numbers $(i, j)$ in each node denotes coordinate $(i/2^3, j/2^3)$.}
	\label{fig:network2DLASM3bits}
\end{figure*}

\Figure[!htb](topskip=0pt, botskip=0pt, midskip=0pt){s1_e3_m2_d2_re_index3}
{The functional graph of 2D-LASM with 6-bit floating-point precision and round quantization, where the length of mantissa fraction is 3.\label{fig:6bitprecisionl3m2}}	

\section{Cryptanalysis}
\label{cryptanalysis}

In \cite{yegd:IJBC:2018}, various aspects of IEAIE were analyzed to conclude that it owns superior security performance. However, we try to demonstrate that all the arguments are groundless.

\subsection{Some security defects of IEAIE}
\label{defects}

In \cite{Alvarez:IJBC:2006,Li:RCES:JSS2008}, some rules and suggestions for designing secure and efficient image encryption schemes were concluded. Some concrete steps for evaluating security performances of a chaotic image encryption schemes were given in \cite{ozkaynak:review:ND18}. Unfortunately, IEAIE did not follow the lessons summarized in \cite{Alvarez:IJBC:2006,Li:RCES:JSS2008,ozkaynak:review:ND18}. To attract the attention of designers of image encryption schemes on cryptanalysis, we check the security of
every aspect of IEAIE and its test given in \cite{yegd:IJBC:2018} as follows.

\begin{figure*}[!htb]
	\centering
	\begin{minipage}[b]{\threeimagewidth}
		\centering
		\includegraphics[width=\threeimagewidth]{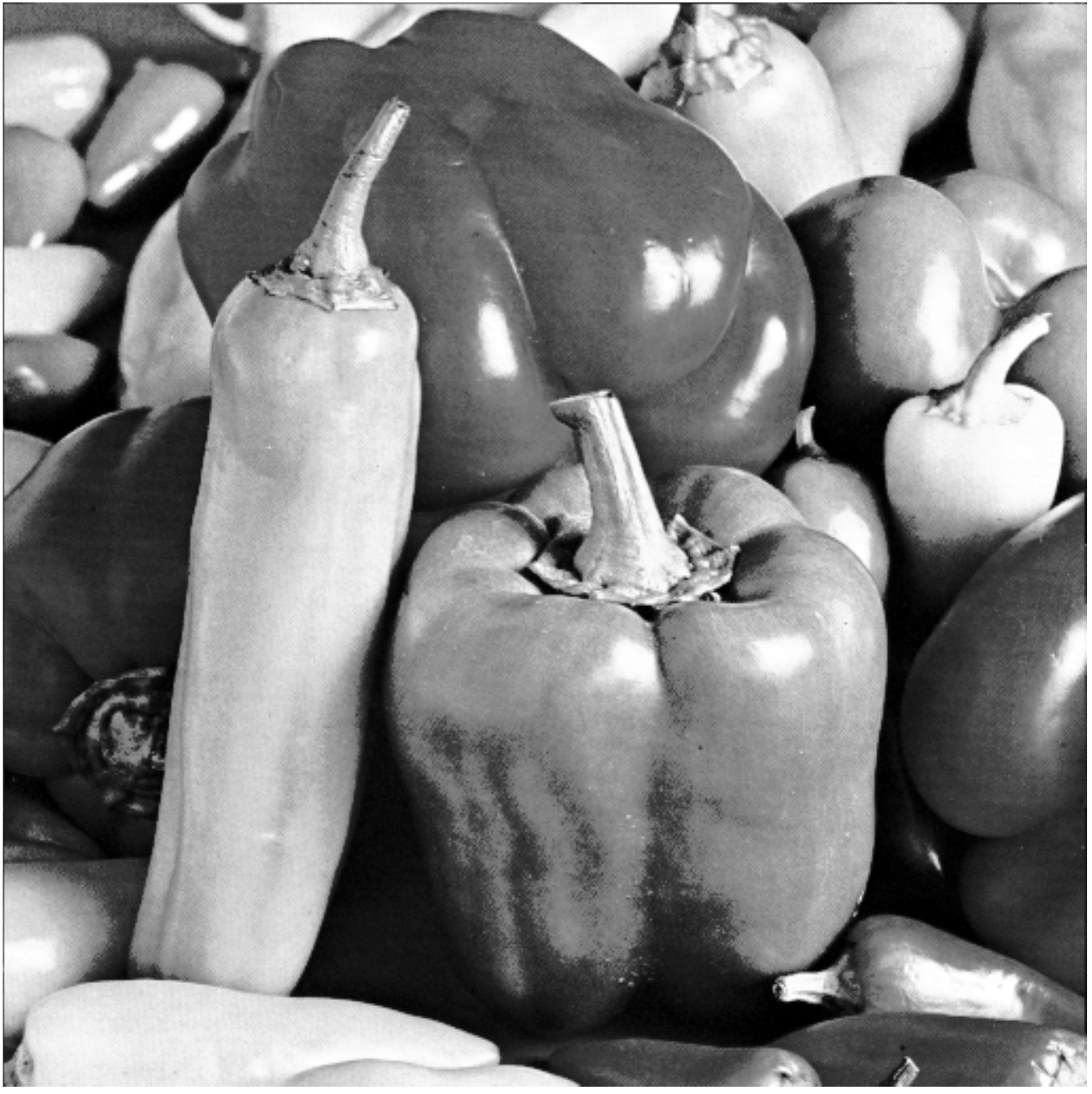}
		a)
	\end{minipage}
	\begin{minipage}[b]{\threeimagewidth}
		\centering
		\includegraphics[width=\threeimagewidth]{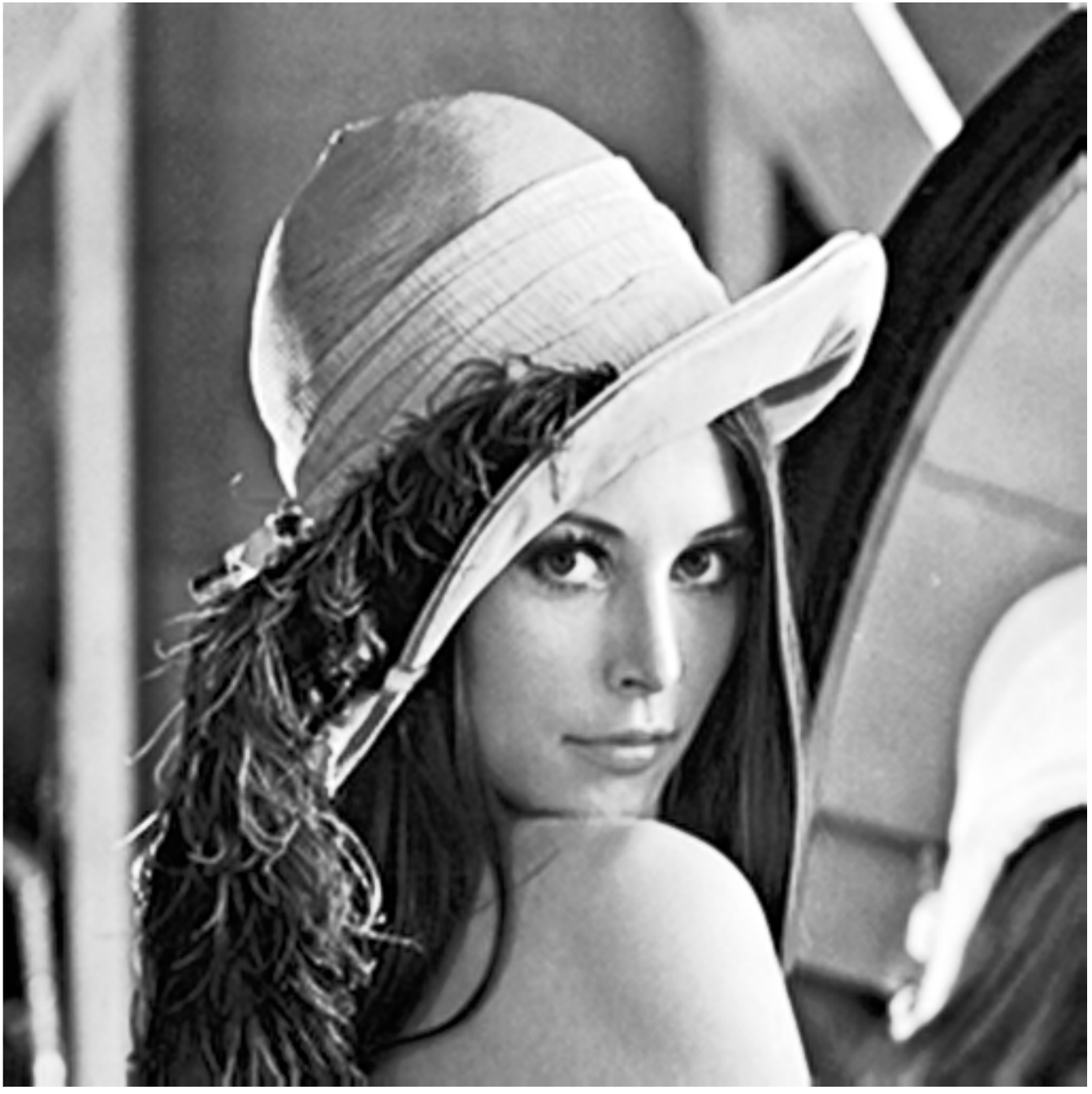}
		b)
	\end{minipage}
	\begin{minipage}[b]{\threeimagewidth}	
		\centering	
		\includegraphics[width=\threeimagewidth]{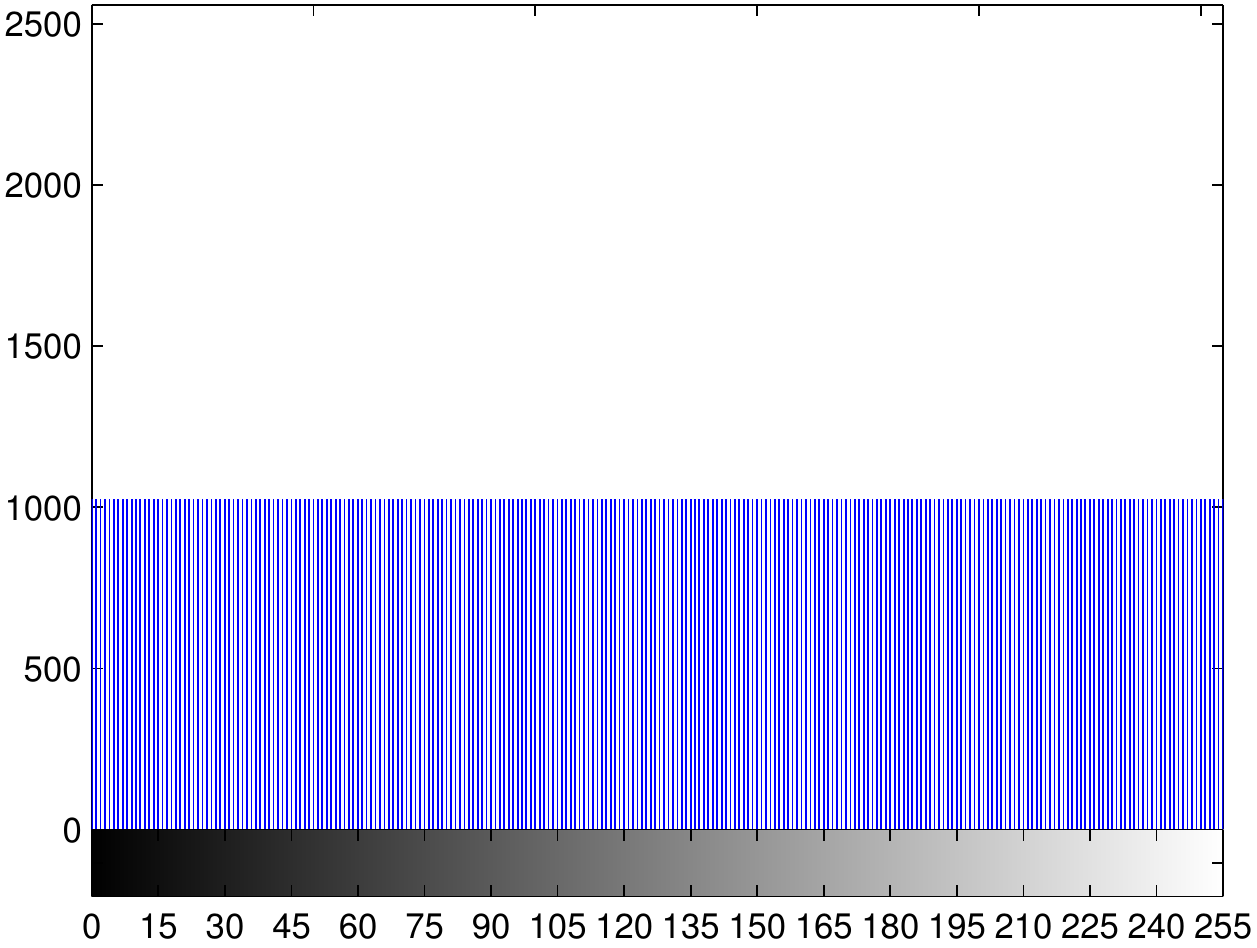}
		c)
	\end{minipage}
	\caption{Two images with the same flat histogram: a) ``Peppers''; b) ``Lenna''; c) histogram of the images shown in Fig.~\ref{fig:Histogram}a), b).}
	\label{fig:Histogram}
\end{figure*}

\begin{itemize}[topsep=0pt,parsep=0pt,partopsep=0pt,leftmargin=10pt,labelwidth=6pt,labelsep=4pt]
\item \textbf{Underlying chaotic map}:

In \cite{Hua:sine:IS2016}, a new two-dimensional chaotic map 2D-LASM was constructed by `adjusting' Logistic map and Sine map with three strategies: cascading output of the former as the input of the latter; extending dimension of the phase plane from 1D to 2D; adopting one more multiplication variable with a constant delay parameter. It was proved that 2D-LASM can demonstrate much more complex chaotic behaviors than the two original 1D maps \cite{hua:Sine:TIE2018}. As shown in Fig.~\ref{fig:network2DLASM3bits}, any orbit will definitely enter a cycle after a transient process. Rigorous theoretical analyses given in \cite{li2017networkChaos} prove that the functional graph of any digital chaotic map is highly correlated with that in a domain with arithmetic precision as small as 3. As shown in Figs.~\ref{fig:network2DLASM3bits}, \ref{fig:6bitprecisionl3m2}, the cycle length of the functional graph of 2D-LASM is very small for either arithmetic format. In \cite{yegd:IJBC:2018}, it was stated that ``previously iterated values were discarded to avoid transient effects''. Actually, the differences between neighboring states change exponentially along an orbit of any iterated map, which is different from the case for chaotic flow. The real purpose of discarding some initial iterated values is to avoid recovering the control parameters of the corresponding chaotic map from them, which is demonstrated in \cite{Li:logistic:ND2014}.
As shown in \cite{li2017networkChaos}, some cycles of short period (even self-loop) always exist no matter which enhancement method is
adopted, e.g. increasing the arithmetic precision, perturbing states, perturbing the control parameters, switching among multiple chaotic maps, and cascading among multiple chaotic maps. If the initial state is located in a small-scale connected component or a cycle of short period in the functional graph of the used chaotic map, there are not enough available states (200 states specified in \cite[Sec.~3.1]{yegd:IJBC:2018}) to be discarded. So, an adaptive threshold should be set to avoid this problem. But, it would cost additional computation.
	
\item \textbf{Key sensitivity}:

In \cite{yegd:IJBC:2018}, ``a small change of $10^{-14}$ is shifted in keys $x_0$, $y_0$, $x'_0$, $y'_0$'' to check their influence on the decryption results. Although $10^{-14}$ is small in itself as for our daily lives, the shift may cause a dramatic change of binary presentation of a number. Let's illustrate this problem with arithmetic format binary32 (single-precision floating-point format), where $10^{-14} =( 1.011 0100 0010 0100 1101 1100)_2\cdot 2^{-47}$ (stored as binary string ``$0 01010000\uline{ 011 0100 0010 0100 1101 1100}$'').
As for number $10^{-12} = (1.000 1100 1011 1100 1100 110$ $0)_2\cdot 2^{-40}$ (``$0 01010111 \uline{000 1100 1011 1100 1100 1100}$''),
$10^{-12}-10^{-14} =$$(1.000 1011 0101 0100 1000 0010)_2\cdot 2^{-40}$ (``$0 01010111 \uline{000 1011 0101 0100 1000 0010}$''). It can be calculated that 11 bits among the 23 fraction bits (underlined parts) of the subtracted number are changed. So, the four cases given in \cite[Fig.~4]{yegd:IJBC:2018} are far not enough to convince us anything. Now, we can see that a small decimal number should not be used to measure the change degree of initial condition in a binary computer.

Observing Eq.~\eqref{eq:2DLASM}, one can see that $(x_0, y_0)$ and $(1-x_0, 1-y_0)$ are equivalent if 2D-LASM is implemented in a fixed-point arithmetic domain. Due to the modulo addition and division operation in Eq.~\eqref{eq:getMatrixP}, there may exist even much more equivalent secret keys. Besides these, quantization effects of the digital chaotic map may generate the same iteration orbit for different initial conditions (See Figs.~\ref{fig:network2DLASM3bits}, \ref{fig:6bitprecisionl3m2}). So, the sensitivity of encryption results of IEAIE with respect to the change of its secret key is very weak.

\item \textbf{Key space analysis}:
	
In \cite[Sec. 3.2.1]{yegd:IJBC:2018}, the precision of the secret key of IEAIE is fixed as $10^{-14}$. In digital world, the precision can only be precisely specified by a power of two. If a floating-point number format (binary32 or binary64) is adopted, the distances between neighboring representable numbers are not uniform, which requires setting the length of mantissa fraction, and that of exponent, elaborately \cite{li2017networkChaos}. As shown in Fig.~\ref{fig:network2DLASM3bits} and \cite{hua:Sine:TIE2018}, there exists a number of nonchaotic regions of $(x_0, y_0)$. The initial conditions falling in such regions may compose invalid or weak secret keys. So, we can conclude that the specification of IEAIE seriously violates Rule~5 suggested in \cite{Alvarez:IJBC:2006}, ``The key space $\mathcal{K}$, from which valid keys are to be chosen, should be precisely specified and avoid nonchaotic regions.'' In addition, the computational complexity of checking each secret key and the valid time of the protected plain-image are not considered in \cite{yegd:IJBC:2018}. In all, the statement ``the brute-force attack is impossible to successfully execute'' is unconvincing.
	
\item \textbf{Histogram}:
	
In \cite{yegd:IJBC:2018}, it was emphasized that ``the histogram of the cipher-image should be or near uniform and be different from that of the plain-image after encryption''. In fact, what counted for the security of IEAIE should be the matching degree between secret-key (or plain-image) and the histogram of the corresponding cipher-image.
As shown in\cite{cqli:autoblock:IEEEM18}, an attacker can recover some statistical information of the plain-image by changing the counting objects of the histogram from pixel to bit. In addition, the spatial information of pixels may play a dominant role for the visual effect of the composed image. To show this point, Fig.~\ref{fig:Histogram} presents two $512\times 512$ images with the same flat (exactly uniform) histogram, whose number of pixels for each tonal value is $\frac{512 \cdot 512}{256}=1024$. Figure~\ref{fig:Histogram_e} gives the encryption results of the two images shown in Fig.~\ref{fig:Histogram} with the position permutation-only scheme HCIE cryptanalyzed in \cite{Cqli:hierarchical:SP2016}. Although histograms of the two encrypted images kept unchanged, the scheme is secure enough for some application scenario, e.g. surveillance and protection of pay-TV from illegal users. Anyway, the three histograms calculated in terms of pixel shown in
\cite[Fig. 5]{yegd:IJBC:2018} are not far enough to prove ``the proposed algorithm has a good ability to frustrate the attack'' based on the histogram.

\begin{figure*}[!htb]
	\centering
	\begin{minipage}[b]{\threeimagewidth}
		\centering
		\includegraphics[width=\threeimagewidth]{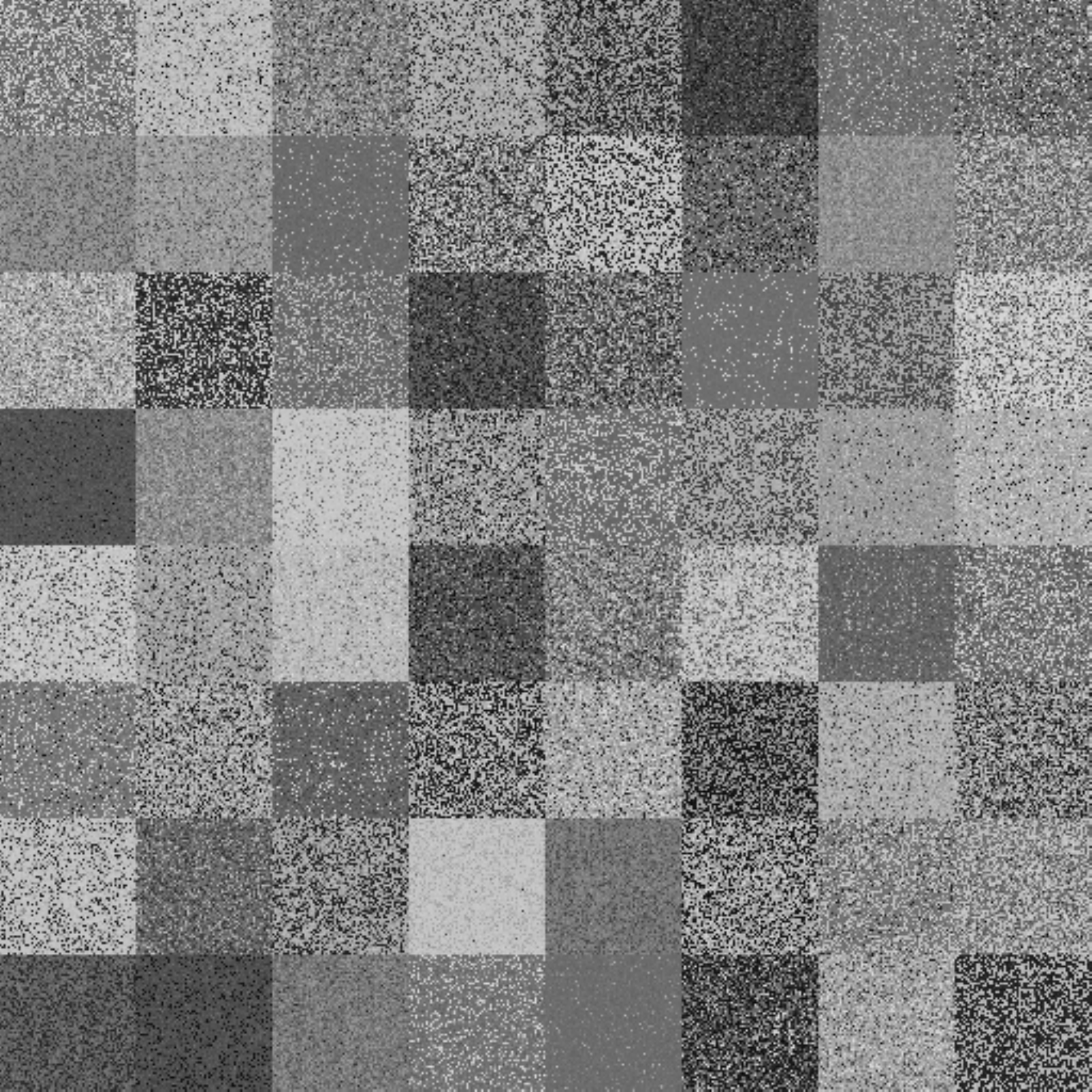}
		a)
	\end{minipage}
	\begin{minipage}[b]{\threeimagewidth}
		\centering
		\includegraphics[width=\threeimagewidth]{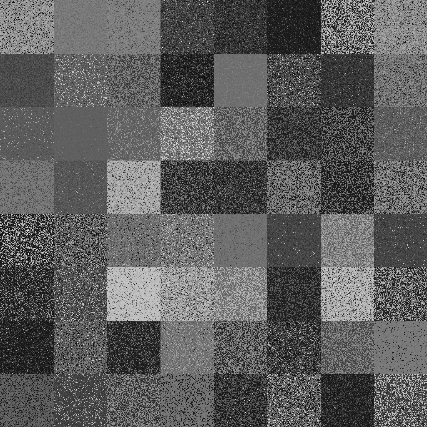}
		b)
	\end{minipage}
	\begin{minipage}[b]{\threeimagewidth}	
		\centering	
		\includegraphics[width=\threeimagewidth]{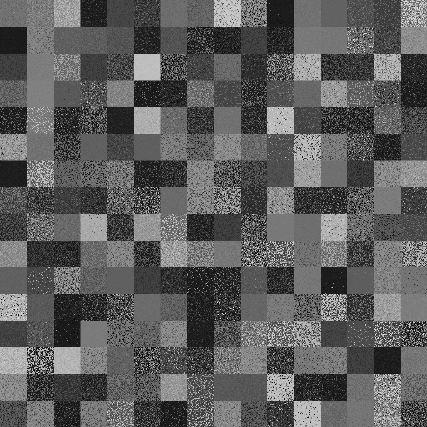}
		c)
	\end{minipage}
	\caption{Three cipher-images encrypted by HCIE: a) ``Peppers'' with $64\times 64$ blocks; b) ``Lenna'' with $64\times 64$ blocks; c) ``Lenna'' with $32\times 32$ blocks}
	\label{fig:Histogram_e}
\end{figure*}
	
\item \textbf{Variance of histogram}:
	
To further measure the uniformity degree of a cipher-image, the variance of its histogram was calculated in \cite{yegd:IJBC:2018}. Actually, the variance of a histogram cannot measure the number of different possible histograms generated by a tested encryption scheme. For example, the variances of two histograms ``2, 2, 3, 4, 7'' and ``2, 2, 3, 5, 6'' are different, but their number of different combinations are the same. But, the histogram variances of four cipher-images given in \cite[Table~4]{yegd:IJBC:2018} are far not sufficient to demonstrate existence of any rule. In addition, even some insecure encryption schemes can also make the obtained cipher-image own very low variance of histogram \cite{Uhl:chaos:TIFS2018}. Moreover, visual security indexes of the three cipher-images shown in Fig.~\ref{fig:Histogram_e} are different, but the variances of the histogram of them are fixed to zero. In all, the statement ``a lower variance represents higher uniformity'' in \cite{yegd:IJBC:2018} is not right.
	
\item \textbf{Information entropy}:
	
Information entropy is a quantitative metric measuring the disorder or randomness in a closed system. From Eq.~\eqref{eq:entropy}, one can see that the entropy value of a message kept unchanged with respect to the following two kinds of changes: 1) permuting the position of every element within the message; 2) changing the elements of a given value as another one that does not exist in the message (if there is)  \cite{Madiman:entropy:RSA20385}. In each case, the changes compose a bijection between specific domain
and the corresponding codomain (See Fact~\ref{entropyProperty}). In all, there are a huge number of different images owning the same information entropy as a given image when its size is relatively large. For example, the five different images shown in Fig.~\ref{fig:Histogram}, \ref{fig:Histogram_e} share the same value of information entropy. 
Embedding Eq.~\eqref{equivalentPermut} into Eq.~(\ref{eq:changeDistribution}), one can see that
$H(\mathbf{R}_j )$ is determined by the two matrixes $\mathbf{P}$ and $\mathbf{T}$ for a given plain-image $\mathbf{I}$, where $j\in \{1, 2, \cdots, N\}$. From the definition of $\mathbf{R}_j$
and $\mathbf{T}$, one can deduce that every column of $\mathbf{T}$ should be of fixed value
to assure that the modulo addition in Eq.~(\ref{eq:changeDistribution}) has the same effect on 
$\{\mathbf{R}_j\}_{j=1}^N$, namely every column of $\mathbf{R}$ (the difference between $\mathbf{R}_{j}$ and $\mathbf{R}_{j+1}$), for different plain-images. To satisfy such condition, $N\equiv 0 \pmod{256}$ should hold. Even this, the statement ``the value of the information entropy is very sensitive to the message'' given in \cite[Sec. 2.1]{yegd:IJBC:2018} is still baseless. Note that the tiny differences of entropy given in \cite[Table~7]{yegd:IJBC:2018} are only bounded by 0.01 and the cipher-image of ``Lenna'' encrypted by the analyzed bit-level permutation-only scheme cryptanalyzed in \cite{Cqli:Scramble:IM17} can also reach as high as 7.978.

\begin{fact}
For any function $f$, entropy function $H(\mathbf{X})$ (Eq.~(\ref{eq:entropy})) satisfies that $H(f(\mathbf{X}))\leq H(\mathbf{X})$ and the equality holds if and only if $f$ is a bijection.
\label{entropyProperty}
\end{fact}

\item \textbf{Plaintext sensitivity}:
	
Plaintext sensitivity is very important for high-strength image encryption schemes as a plain-image and its slightly modified version (embedded by a watermark or some hiding messages) are often encrypted at the same time. If the used encryption scheme does not satisfy the sensitivity requirement, leakage of the cipher-image corresponding to one of the two similar plain-image may disclose the visual information of the other. In the field of image security, two metrics
UACI (unified averaged changed intensity) and NPCR (number of pixels changing rate) are widely used to measure plaintext sensitivity. Unfortunately, the validity of the two metrics has been questioned in \cite{Uhl:chaos:TIFS2018} by statistical information of the outputs of some insecure encryption schemes. Here, we emphasize that the internal structure of IEAIE cannot perform well to achieve the expected plaintext sensitivity. Observing the encryption procedure of IEAIE, one can see that all involved operations can make every operated bit `run' from the least significant bit (LSB) to the most significant bit (MSB), not the opposite order. Concretely, the change of a bit in the $i$-th bit-plane (counted from the LSB to MSB) can only influence the bits in the $i\sim 8$-th ones. So, the influence scope of every bit of the plaintext on the corresponding cipher-text is dramatically different. No matter how many round numbers are repeated, this problem remains to exist \cite{Cqli:Fridrich:SP2017}. The designers of IEAIE claimed that ``the keystreams are different with respect to different plain-images'' based on the assumption of high sensitivity of information entropy on change of the plain-image. However, as we have explained above, this assumption is not correct. In all, the statement ``a slight change in the plain-image leads to a completely different cipher-image'' in \cite[Sec.~3.2.2]{yegd:IJBC:2018} is incorrect.
	
\begin{figure*}[!htb]
\centering
\begin{minipage}[b]{2\threeimagewidth}
		\centering
\includegraphics[width=2\threeimagewidth]{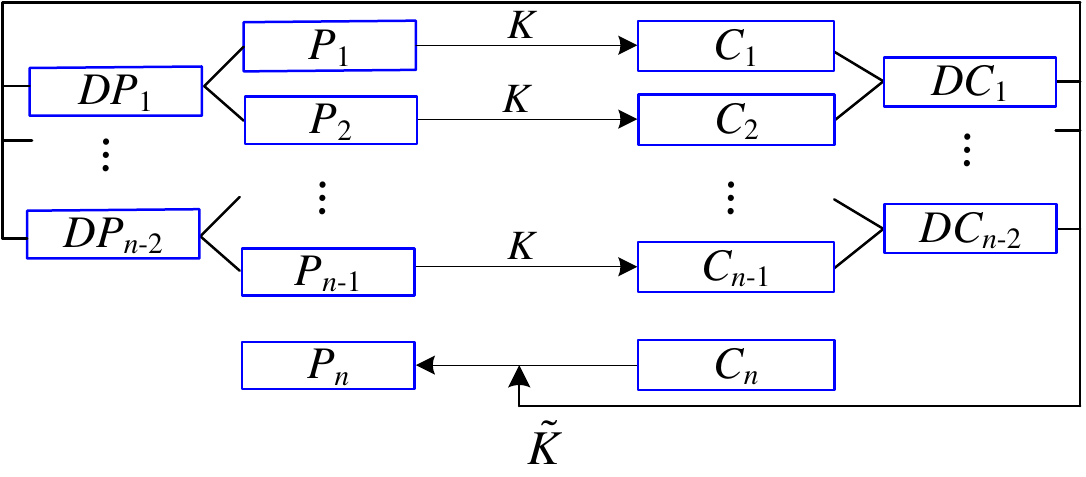}
\end{minipage}
\caption{The model of differential attack.}
\label{fig:difference}
\end{figure*}
	
\item \textbf{Coefficient correlation}:
	
Just like most chaos-based image encryption schemes, \cite{yegd:IJBC:2018} adopted the 
coefficient correlation of neighboring pixels of cipher-images encrypted by IEAIE to demonstrate its good security performance. As mentioned in \cite[Fig.~3]{Uhl:chaos:TIFS2018}, there is ``no clear (statistical) decision criterion for passing this test''. Furthermore, three insecure encryption schemes deliberately constructed in \cite{Uhl:chaos:TIFS2018} can perform very well in terms of fulfilling the metric. Actually, this metric can be calculated only from image encryption schemes working in the spatial domain. Reasonable security index of image data should consider the characteristics of the human visual system and the distribution of compressing coefficients of image data \cite{xt:VisualSecurity:TIFS16}.
	
\item \textbf{Efficiency analysis}:
	
The authors of \cite{yegd:IJBC:2018} claimed that IEAIE is suitable for real-time secure communication by comparing it with the image encryption scheme proposed in \cite{Hua:sine:IS2016}. In fact, the fast running speed of IEAIE comes from less computation operations, namely the obtained efficiency is built on sacrificing security instead of better structure.
In Eqs.~\eqref{eq:getMatrixP}, \eqref{convert2}, \eqref{convert}, \eqref{parameterD}, IEAIE uses integer conversion functions
following the general form
\begin{equation}
f_n(x)=f\left(10^m \cdot x\right) \bmod D,
\label{multiply}
\end{equation}
where $m$ and $D$ are positive integers, $f(x)$ is a quantization function, e.g. ceil and round.
In processors, multiplication by a constant is implemented using a sequence of bit-wise shift and addition operations, e.g.
\[
g(x)=((x\lll 2) + x) \lll 1,
\]	
where
\[x\lll s=\sum_{i=0}^{L-s-1}\left(x_i\cdot 2^{i+s}\right),\]
$x=\sum_{i=0}^{L-1}x_i\cdot 2^i$, and $L$ is the arithmetic precision. So the computational complexity of the conversion (\ref{multiply}) is proportional to $m$ \cite{cqli:autoblock:IEEEM18}.
In Eqs.~\eqref{eq:getMatrixP}, \eqref{convert2}, \eqref{convert}, \eqref{parameterD}, $m$ is set as 7 or 14. Only $\lceil \log_2 D \rceil$ bits are useful for IEAIE, the computation spent on generating the other $m\lceil \log_2(10)  \rceil-\lceil \log_2 D \rceil$ bits are wasted. Taking Eq.~\eqref{convert} as an example, the utilization
percentage of the computation cost on iterating 2D-LASM \eqref{eq:2DLASM} is only $\frac{\lceil \log_2 D \rceil}{m\lceil \log_2(10)\rceil }=\frac{\lceil \log_2 256 \rceil}{14\cdot \lceil \log_2(10)}=\frac{1}{7}$.
In addition, the test on speed analysis in \cite{yegd:IJBC:2018} was performed in the idea laboratory environment instead of resource-limiting real environments.
\end{itemize}

\subsection{Differential cryptanalysis}

As shown in Fig.~\ref{fig:difference}, an attacker can arbitrarily choose some plaintexts, $P_1, P_2, \cdots, P_{n-1}$, and
the corresponding ciphertexts, $C_1, C_2, \cdots, C_{n-1}$, encrypted by the same secret key $K$ in the scenario of chosen-plaintext attack. As for differences between plaintexts $P_i$ and $P_{i+1}$, $\mathit{DP}_1, \mathit{DP}_2, \cdots, \mathit{DP}_{n-1}$, one can observe the corresponding differences between ciphertexts $C_i$ and $C_{i+1}$, $\mathit{DC}_1, \mathit{DC}_2, \cdots, \mathit{DC}_{n-1}$. The differences are defined in terms of an invertible operation used in the encryption scheme, e.g. bitwise OR and
modulo subtraction. So differential cryptanalysis can be considered as a chosen-plaintext attack on a weakened version of
the analyzed encryption scheme for some differences selected from ${n\choose 2}=n(n-1)/2$ possible ones.
In the broadest sense, \textit{differential cryptanalysis} is a cryptanalytic method studying how particular differences in {plaintext} pairs affect the resultant differences, which is also called \textit{differential}, of the corresponding ciphertext pairs. Considering the public structure of the analyzed encryption scheme, some basic parts can be deliberately canceled and the remaining part can be broken with much less resources. By repeating the process, the equivalent secret key of the whole encryption scheme
$\tilde{K}$ can be recovered, which is then used to decrypt another ciphertext $C_{n}$, encrypted by the same secret key.

From the above introduction of the chosen-plaintext attack, one can see that the attack model relies on repeating usage of the secret key. So the designers of IEAIE use the information entropy of the plain-image to ``affect the usage of the keystream and frustrate the chosen-plaintext and known-plaintext attacks'' in \cite[Sec. 2.2]{yegd:IJBC:2018}. However, based on the analysis on the insensitivity of information entropy in above sub-section, it is very easy to construct some plain-images possessing the same keystream during the encryption process of IEAIE. Observing the encryption procedure of IEAIE, one can see that its real operations are solely determined by $N+1$ parameters, $s$ and $\{d_j\}_{j=1}^N$. Note that even when two plain-images generate different entropy values in the encryption process, their corresponding key steams are still maybe the same due to the following reasons:
1)  the modulo addition and division in Eq.~(\ref{eq:update}) may make different sets of $(x_0, y_0, x'_0, y'_0, s)$ result in the same value of $(\bar{x}_0, \bar{y}_0)$; 2) the quantization error of calculating $\log_2(\cdot)$ in computer may make different combinations of $\{\phi_i \}_{i=0}^{2^8-1}$ generate the same value of $H(X)$; 3) Eq.~\eqref{parameterD} only extract $\lceil \log_2(N) \rceil $ bits of intermediate computing result of  $\mathbf{R}_j$, and may output the same value of 
$d_j$ for different inputs of $\mathbf{R}_j$. Once the dependability mechanism of the key stream of IEAIE on the plain-image is concealed, the structure of Eq.~\eqref{eq:diffusion} becomes the same as that of the main function of the image encryption scheme cryptanalyzed in \cite{cqli:autoblock:IEEEM18}. Then, the differential cryptanalysis on IEAIE can be performed similarly.

Assume two plain-images $\mathbf{I}$ and $\mathbf{I}'$ own the same set of $s$ and $\{d_j\}_{j=1}^N$ in the encryption process of IEAIE. As for their difference in terms of modulo subtraction $\Delta\mathbf{I}$, IEAIE is degenerated to
\begin{equation}
\begin{cases}
\Delta\mathbf{R}( \mathbf{P}(i, j ))=\Delta\mathbf{I}(i, j),\\
\Delta\mathbf{C}(i, j)=(\Delta\mathbf{R}(i, j)+ d_j \cdot \Delta\mathbf{C}(i, j-1)) \bmod 256,
\end{cases}
\label{differentialEncrypt}
\end{equation}
where $i=1\sim M$, $j=1\sim N$, $\Delta\mathbf{C}$ is the difference of the corresponding cipher-images of $\mathbf{I}$ and $\mathbf{I}'$ in terms of the operator
(some components in Eq.~(\ref{eq:diffusion}) are eliminated by the modulo subtraction), and $\Delta\mathbf{C}(i, 0)\equiv 0$. Observing Eq.~(\ref{differentialEncrypt}), one can
assure that 
\begin{equation}
\mathbf{P}(i^*, j^*)=(i^{**}, j^{**})
\label{RecoverPermute}
\end{equation}
if $\Delta\mathbf{I}$ has only one non-zero element at entry $(i^*, j^*)$, 
where $(i^{**}, j^{**})$ is the entry of the first non-zero element in differential $\Delta_\mathrm{C}$ (counted in the scan order).

Based on the above analysis, the differential cryptanalysis on one round version of IEAIE can be described as follows.
\begin{itemize}
\item \textit{Step 1}: Choose two plain-images of size $M\times N$, $\mathbf{I}$ and $\mathbf{I}'$, satisfying 
\begin{equation*}
	\begin{cases}
	\mathbf{I}(i^*, j^*)=a, \\
	\mathbf{I}'(i^*, j^*)=b, \\
	\mathbf{I}(i, j)=c, \\
	\mathbf{I}'(i, j)=c,
	\end{cases}
\end{equation*}
where $a, b, c$ are non-negative integers and $\#\{a, b, c\}=3$, $(i, j)\in \{(1, 1),(1, 2),\cdots (M, N)\}\setminus{(i^*, j^*)}$, and $\#\{\cdot\}$ denotes the cardinality of a set. Note that $(i^*, j^*)$ is initialized as $(1, 1)$.

\item \textit{Step 2}: Let $\mathbf{I}$ and $\mathbf{I}'$ pass through the encryption process of IEAIE
with an unknown secret key and obtain the corresponding cipher-images, $\mathbf{C}$ and $\mathbf{C}'$.

\item \textit{Step 3}: Get the value of $\mathbf{P}(i^*, j^*)$ via Eq.~(\ref{RecoverPermute}).

\item \textit{Step 4}: Repeat the above procedure for $(i^*, j^*)=(1, 2), (1, 3) \sim(M, N-1)$ (selected in the scan order of a matrix of size $M\times N$) and
get the value of  $\mathbf{P}(i^*, j^*)$. The value of $\mathbf{P}(M, N)$ can be  identified as the sole unused location.

\item \textit{Step 5}: Recover $\mathbf{R}$ by Eq.~\eqref{equivalentPermut} and calculate
\[\mathbf{D}(i, j) = ( \mathbf{C}(i, j)-\mathbf{R}(i, j)-d\cdot \mathbf{C}(i, j-1)) \bmod 256\]
for $i=1\sim M$ and $j=1\sim N$. 
\end{itemize}

\begin{figure*}[!htb]
	\centering
	\begin{minipage}[b]{1.1\imagewidth}
		\centering
		\includegraphics[width=1.1\imagewidth]{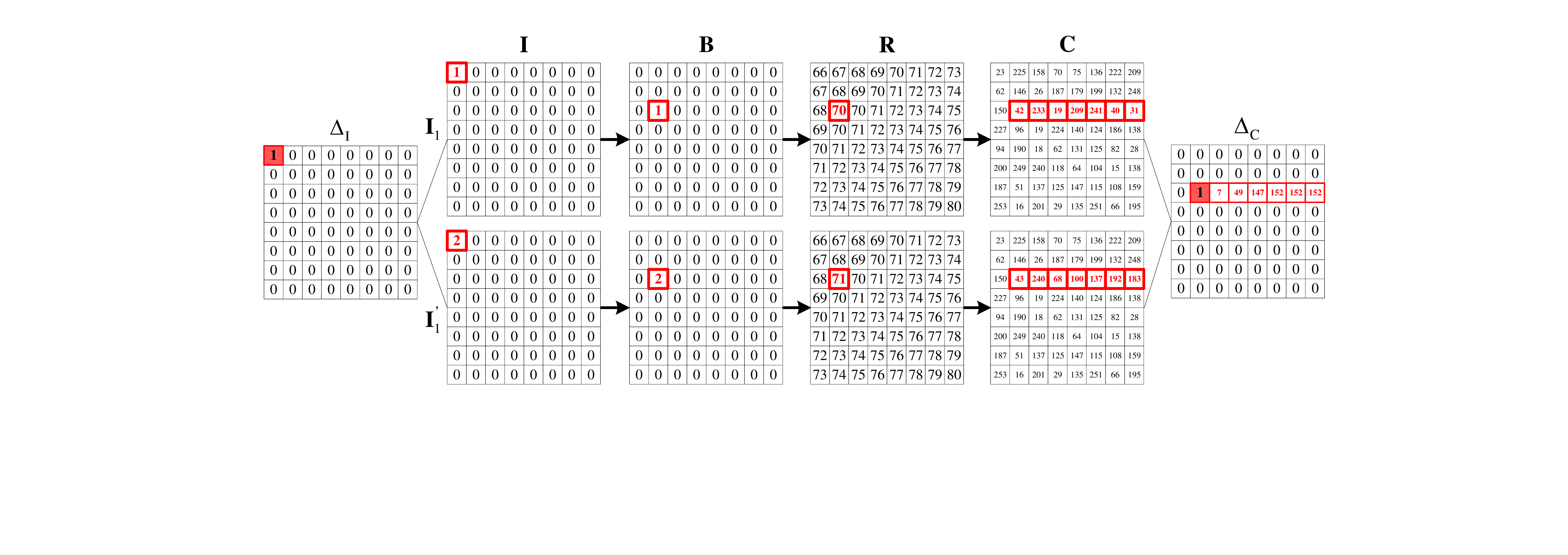}
	\end{minipage}
	\caption{The process of revealing the permutation procedure of IEAIE with a differential attack.}
	\label{fig:example}
\end{figure*}

Observing Eq.~\eqref{eq:diffusion}, one can see that
$\mathbf{D}=\{\mathbf{D}(i, j)\}_{i=1, j=1}^{M, N}$ can work as the equivalent version of the secret key
for decryption on the \textit{diffusion encryption} part. In all, two matrixes $\mathbf{P}$ and $\mathbf{D}$ can work as the equivalent secret key of IEAIE. 

A number of experiment were performed to verify performance of the above attacking steps. A concrete case of revealing the permutation relationship of IEAIE on a plain-image of size $8\times 8$ is shown in Fig.~\ref{fig:example}, where $x_0=0.0056$, $y_0=0.3678$, $x_0'=0.6229$, and $y_0'=0.7676$, and $\mu=0.8116$. In this case, the same set of $s$ and $\{d_j\}_{j=1}^8$ are generated for the two toy plain-images due to the quantization effect.
From Fig.~\ref{fig:example}, we can see that the new permuted location of the sole non-zero element at entry $(1, 1)$ in the differential plain-image
can be observed by searching for the first different elements of the two cipher-images.

In \cite{yegd:IJBC:2018}, two rounds of the basic operations are suggested. Here, we skip the cryptanalysis of the full version of IEAIE based on the following considerations: 1) existence of the security defects presented in the above sub-section is not related with the round number; 2) cryptanalysis of the two rounds of IEAIE involves very complex deduction and presentation; 3) the reported security defects of IEAIE are far enough to demonstrate that it cannot be fixed by simple modifications.

\section{Conclusion}

This paper analyzed the security of a chaotic image encryption algorithm based on information entropy, IEAIE. The claimed superiorities of its structure are analyzed in detail and are found incorrect. Furthermore, every used security metric is incapable to testify real security performance. To design a secure and efficient multimedia encryption scheme, the related critical factors, e.g. the special properties of multimedia data, the concrete application scenario with specified constraints, computation load, should be considered comprehensively. Much cryptanalytic works need to be done to bridge the  gap between the field of nonlinear dynamics and that of modern cryptography.


\bibliographystyle{IEEEtran_doi}
\bibliography{ye_IJBC}

\vspace{18cm}

\EOD
\end{document}